\documentclass[pdflatex,sn-mathphys-num]{sn-jnl}% Math and Physical Sciences Numbered Reference Style 
\usepackage{graphicx}%
\usepackage{subcaption}
\usepackage{multirow}%
\usepackage{amsmath,amssymb,amsfonts}%
\usepackage{amsthm}%
\usepackage{mathrsfs}%
\usepackage[title]{appendix}%
\usepackage{xcolor}%
\usepackage{textcomp}%
\usepackage{manyfoot}%
\usepackage{booktabs}%
\usepackage{algorithm}%
\usepackage{algorithmicx}%
\usepackage{algpseudocode}%
\usepackage{listings}%
\usepackage{subcaption} % Required for subfigures
\usepackage{tcolorbox}
\usepackage{lipsum}
\usepackage{xcolor}
\theoremstyle{thmstyleone}%
\newcommand{\mypar}[1]{\vspace{1pt}\noindent\textbf{#1.}}
\newcommand{\mypartwo}[1]{\vspace{1pt}\noindent\textit{#1.}}

\hypersetup{
	pdftoolbar=true,        % show Acrobat?s toolbar?
	pdfmenubar=true,        % show Acrobat?s menu?
	pdffitwindow=false,     % window fit to page when opened
	pdfstartview={FitH},    % fits the width of the page to the window
	pdftitle={On the Potential of Large Language Models Solve to Semantics-Aware Process Mining Tasks},    % title
	pdfauthor={Adrian Rebmann, Fabian David Schmidt, Goran Glavas, and Han van der Aa},     % author
	pdfsubject={},   % subject of the document
	pdfcreator={},   % creator of the document
	pdfproducer={}, % producer of the document
	pdfkeywords={}, % list of keywords
	pdfnewwindow=true,      % links in new window
	colorlinks=true,       % false: boxed links; true: colored links
	linkcolor=blue,          % color of internal links (change box color
	citecolor=blue,        % color of links to bibliography
	filecolor=blue,      % color of file links
	urlcolor=blue           % color of external links
}

\begin{document}

\title[On the Potential of Large Language Models Solve to Semantics-Aware Process Mining Tasks]{On the Potential of Large Language Models to Solve Semantics-Aware Process Mining Tasks}


\author*[1]{\fnm{Adrian} \sur{Rebmann}}\email{adrian.rebmann@sap.com}

\author[2]{\fnm{Fabian David} \sur{Schmidt}}\email{fabian.schmidt@uni-wuerzburg.de}

\author[2]{\fnm{Goran} \sur{Glava{\v{s}}}}\email{goran.glavas@uni-wuerzburg.de}

\author[3]{\fnm{Han} \spfx{van der} \sur{Aa}}\email{han.van.der.aa@univie.ac.at}

\affil*[1]{\orgdiv{SAP Signavio}, \orgname{SAP SE}, \orgaddress{\street{Dietmar-Hopp-Allee 16}, \city{Walldorf}, \postcode{69190}, \state{Baden-Württemberg}, \country{Germany}}}

\affil[2]{\orgdiv{Center for Artificial
Intelligence and Data Science}, \orgname{University of Würzburg}, \orgaddress{\street{Sanderring 2}, \city{Würzburg}, \postcode{97070}, \state{Bavaria}, \country{Germany}}}

\affil[3]{\orgdiv{Faculty of Computer Science}, \orgname{University of Vienna}, \orgaddress{\street{Währinger Str.\ 29}, \city{Vienna}, \postcode{1090}, \country{Austria}}}

\abstract{Large language models (LLMs) have shown to be valuable tools for tackling  process mining tasks. Existing studies report on their capability to support various data-driven process analyses and even, to some extent, that they are able to reason about how processes work.
This reasoning ability suggests that there is potential for LLMs to tackle semantics-aware process mining tasks, which are tasks that rely on an understanding of the meaning of activities and their relationships. Examples of these include process discovery, where the meaning of activities can indicate their dependency, whereas in anomaly detection the meaning can be used to recognize process behavior that is abnormal.
In this paper, we systematically explore the capabilities of LLMs for such tasks. Unlike prior work, which largely evaluates LLMs in their default state, we investigate their utility through both in-context learning and supervised fine-tuning. Concretely, we define five process mining tasks requiring semantic understanding and provide extensive benchmarking datasets for evaluation. Our experiments reveal that while LLMs struggle with challenging process mining tasks when used out of the box or with minimal in-context examples, they achieve strong performance when fine-tuned for these tasks across a broad range of process types and industries.
}




\maketitle

\section{Introduction}
\label{sec:introduction}
%Context
Process mining focuses on analyzing event data from organizational processes to uncover actionable insights about their actual execution. Common process mining tasks include the discovery of process models, recognizing abnormal behavior, and making predictions about the future of ongoing cases. The majority of techniques that have been developed to tackle these tasks are based on the analysis of frequencies, e.g., how often a certain activity $A$ is followed by a activity $B$ in an event log. 
While such techniques have been proven to be successful, recent works have recognized the potential of incorporating considerations of activity semantics, i.e., the meaning of activities, when conducting process mining tasks such as discovery, anomaly detection, and prediction~\cite{van2021natural,caspary2023does,norouzifar2024bridging}.

Given that such semantics-aware analyses exploit the textual labels associated with events, the advent of large language models (LLMs) is a particularly promising development, due to their impressive capabilities when it comes to handling natural language. 
Language models have proven useful for semantic anomaly detection~\cite{caspary2023does, busch2024xsemad} and preliminary explorations of LLMs for interacting with event data show potential~\cite{berti2023abstractions, jessen2023chit, torres2024mapping}. 
However, existing works either focus on evaluating one smaller language model that was fine-tuned on semantic anomaly detection or only use LLMs out of the box to solve general process analysis questions instead of clearly defined process mining tasks. Consequently, a systematic and in-depth study of the  application of LLMs to semantics-aware process mining tasks remains absent.
This research gap largely follows from the lack of well-defined natural language processing (NLP) tasks that effectively conceptualize the capability to perform semantics-aware process mining tasks, along with the  corresponding benchmarking datasets that are necessary for structured evaluations. Addressing these issues is crucial for enabling robust and comparative assessments of LLMs for tackling process mining tasks. 

%Contributions / Results
Recognizing this, our work makes the following contributions:
\begin{itemize}
\item We define five tasks designed to evaluate the capabilities of LLMs in semantics-aware process mining. These tasks focus on (forms of) anomaly detection, next activity prediction, and process discovery. 
These tasks are motivated by the potential benefits they offer in enhancing approaches to their non-semantic counterparts.
Anomaly detection can be improved by identifying undesired process behavior based on activity meaning, such as detecting that a delivery is created after a canceled purchase order. Similarly, next activity prediction benefits by narrowing options to semantically valid choices, e.g., discarding a check request if already approved. Process discovery can leverage activity meaning to, among others, handle event log incompleteness, inferring parallel executions even if not explicitly recorded. For example, in order fulfillment, packaging and invoicing might occur simultaneously, but a frequency-based approach could misrepresent them as sequential.

\item We provide benchmarking datasets for each of the proposed tasks, enabling rigorous quantitative evaluations of the performance of LLMs for semantics-aware process mining. The foundation of these datasets is provided by a corpus of process behavior that we derived from the largest publicly available collection of process models.

\item We conduct an experimental evaluation of several open-source LLMs on the proposed tasks and their corresponding benchmarking datasets. The evaluation includes a comparison of LLMs in both in-context learning and fine-tuning settings, alongside discriminative encoder-based language models.
\end{itemize}
In this manner, our work provides the first systematic evaluation of (L)LMs on semantics-aware process mining tasks.
Our experiments reveal that LLMs struggle with these tasks when used out of the box or with minimal in-context examples, but they achieve strong performance when fine-tuned for these tasks.

This paper is an extended and revised version of our earlier work on evaluating the ability of LLMs to solve semantics-aware process mining tasks, published as part of the proceedings of the International Conference on Process Mining 2024~\cite{rebmann2024evaluating}. 
This works extends the conference paper in the following ways.
First, we broadened the scope of our work by introducing two new semantics-aware process mining tasks. These tasks fundamentally differ from the three previously introduced ones, since they focus on generation (of directly-follows relations and process trees), whereas the previous tasks were all classification-based.
For both new tasks, we established and publish a benchmarking dataset to assess and compare the performance of LLMs when solving them. We use these datasets to extend our experimental evaluation, shedding light on the ability of open source LLMs to solve process-oriented generation tasks.
Beyond these conceptual and experimental extensions, we provide more detailed statistics about the process models that our benchmarking datasets are based on. Furthermore, we thoroughly revised and extended the related work section, highlighting recent advancements in evaluating LLMs for data-driven process analysis tasks.

The remainder of this paper is structured as follows. \autoref{sec:preliminaries} introduces preliminary definitions that form the basis for the semantics-aware process mining tasks that we define in \autoref{sec:tasks}. \autoref{sec:datasets} describes the publicly available process behavior corpus and benchmarking datasets that we have established. \autoref{sec:llms} describes the different ways in which we use and adapt LLMs to tackle the proposed tasks.
\autoref{sec:setup} reports on the setup that we used to conduct our evaluation experiments, of which the results are presented in \autoref{sec:results}. Finally, \autoref{sec:related} discusses related work, before we conclude in \autoref{sec:conclusion}.

\section{Preliminaries}
\label{sec:preliminaries}
In this section, we introduce the preliminaries that are essential for the remainder of the paper, covering event data, process models, directly-follows relations and graphs, footprints, process trees, and eventually-follows relations.

\mypar{Event Data} We adopt a simple event model, focusing on the control-flow of a process. A trace $\sigma$ is a sequence that represents the events that have been recorded for the  execution of a single instance of an organizational process. Such a trace consists of a finite sequence of events with each event as a record of the execution of an activity. We denote this as $\sigma = \langle a_1,...,a_n\rangle$,  with $a_i \in \mathcal{A}$, with $\mathcal{A}$ as the universe of possible activities that can be performed in organizational processes. 
An event log $L$ is a finite multi-set of traces. $A_L \subset \mathcal{A}$ denotes the set of activities that appear in the traces of $L$.

\smallskip
\noindent
For instance, we may have an event log $L_1 = [\sigma_1, \sigma_2, \sigma_3]$, with:

$\sigma_1 = \langle \textit{receive order}, \textit{accept order}, \textit{deliver package} \rangle$

$\sigma_2 = \langle \textit{receive order}, \textit{reject order} \rangle$

$\sigma_3 = \langle \textit{receive order}, \textit{deliver package} \rangle$. 

This then gives $A_{L_1} = \{\textit{receive order}, \textit{accept order}, \textit{reject order}, \textit{deliver package}\}$.

\mypar{Process Models}
We define a process model $M$ as the set of executions that are allowed in a process. 
Each execution $\pi$ is represented as an activity sequence $\pi = \langle a_1,...,a_n \rangle$, with $a_i \in \mathcal{A}$. We use $A_M \subset \mathcal{A}$, to denote the set of activities that appear in the sequences of $M$.

\smallskip
\noindent
For instance, $M_1 = \{\pi_1, \pi_2\}$ with:

$\pi_1 = \langle \textit{receive order}, \textit{accept order}, \textit{deliver package} \rangle$

$\pi_2 = \langle \textit{receive order}, \textit{reject order} \rangle$

\noindent
Then, $A_{M_1} = \{\textit{receive order}, \textit{accept order}, \textit{reject order}, \textit{deliver package}\}$. %Note that, although $A_{M_1}=A_{L_1}$, $L_1$'s trace $\sigma_3$ is not allowed by $M_1$.}

\mypar{Process Trees}
A process tree is a hierarchical representation of a process, with a concise notation. Such a tree has a single root node and its leaves correspond to activities.
Commonly, four operator types are used: $\rightarrow$ (sequence operator), $\times$ (exclusive choice operator), $\wedge$ (parallel operator), and $\circlearrowleft$ (loop operator). We follow the definition of van der Aalst~\cite{van2022foundations}:
Let $O = \{\rightarrow ,\times ,\wedge, \circlearrowleft \}$ be the set of operators and $\tau \not \in \mathcal{A}$ be the \emph{silent activity}. A process tree is defined recursively:

\begin{itemize}
    \item if $a \in \mathcal{A} \cup \{\tau\}$, then $T = a$ is a process tree
    \item if $T_1, T_2, \ldots , T_n$ with $n \geq 1$ are process trees and $\oplus \in \{\rightarrow ,\times ,\wedge \}$, then $T = \oplus (T_1,T_2, \ldots, T_n)$ is a process tree
    \item if $T_1, T_2, \ldots , T_n$ with $n \geq 2$ are process trees and $T = \,{\circlearrowleft }(T_1,T_2, \ldots, T_n)$.
\end{itemize}

\smallskip
\noindent
For instance,  $\rightarrow(\textit{receive order}, \times(\rightarrow(\textit{accept order}, \textit{deliver package}), \textit{reject order}))$ is a tree that captures the behavior defined by the execution sequences of $M_1$, i.e., that, after receiving an order, it is either accepted and then delivered, or it is rejected.

\mypar{Directly-Follows Relations and Graphs}
The directly-follows relation $>$ captures that two activities can immediately follow each other, either in the execution sequences of a process model or the traces of an event log. 
For a process model $M$, $x > y$ if there exists an execution sequence $\pi = \langle a_1,...,a_n \rangle \in M$ for which there exists $a_i = x, a_j = y$ with $j = i +1$. The same applies for a log $L$, where these constraints should hold for at least one trace $\sigma \in L$.
 
We define a directly follows graph (DFG) $D$ as a pair $(A, F)$. $A$ is the set of possible activities in a given process and $F$ is a set of pairs  $(x, y)$ that represent the process' directly-follows relations (or edges in the DFG), i.e., $\{(x,y) \in A \times A \mid x > y $

\noindent
For instance, given $M_1$, we obtain a DFG $D_{M_1} = (A_{M_1}, F_{M_1})$, with:  

\noindent $F_{M_1} = $ 
$\{(\textit{receive order}, \textit{accept order}),(\textit{accept order},\textit{deliver package}),(\textit{receive order}, $ $\textit{reject order})\}$.

\mypar{Directly-Follows Footprints}
Following the definition of van der Aalst~\cite{van2022foundations}, let $D=(A,F)$ be a directly-follows graph. $D$ defines a (directly-follows) \emph{footprint} $fp_D: (A \times A) \rightarrow \{\rightarrow, \leftarrow, ||, \#\}$ such that for all $(x, y) \in A \times A$:

\begin{itemize}
    \item $fp_D((x,y)) = {\rightarrow }$ if $(x,y)\in F$ and $(y,x)\notin F$
    \item $fp_D((x,y)) = {\leftarrow }$ if $(x,y)\notin F$ and $(y,x)\in F$
    \item $fp_D((x,y)) = {\Vert }$ if $(x,y)\in F$ and $(y,x)\in F$
    \item $fp_D((x,y)) = {\#}$ if $(x,y)\notin F$ and $(y,x)\notin F$
\end{itemize}

\noindent
For instance, $fp_{D_{M_1}}$ is:

$fp_{D_{M_1}}((\textit{receive order}, \textit{accept order})) $ = $ \rightarrow$

$fp_{D_{M_1}}((\textit{accept order}, \textit{receive order})) $ = $\leftarrow $

$fp_{D_{M_1}}((\textit{receive order}, \textit{reject order})) $ = $ \rightarrow $

$fp_{D_{M_1}}((\textit{reject order}, \textit{receive order})) $ = $ \leftarrow $

$fp_{D_{M_1}}((\textit{accept order}, \textit{deliver package})) $ = $ \rightarrow $

$fp_{D_{M_1}}((\textit{deliver package}, \textit{accept order})) $ = $ \leftarrow $

$\text{All other pairs in } (A_{M_1} \times A_{M_1}) $ = $ \#$

\mypar{Eventually-Follows Relations} 
The eventually-follows relation $\prec$ is a more relaxed ordering relation than the directly-follows relation described above, focusing on activities that can either directly or indirectly follow each other in the activity sequences of a process model or in the traces of a log. 
For a process model $M$, $x \prec y$ if there exists an execution sequence $\pi = \langle a_1,...,a_n \rangle \in M$ for which there exists $a_i = x, a_j = y$ with $i < j$. The same applies for a log $L$, where these constraints should hold for at least one trace $\sigma \in L$.
We use $\mathit{EF}_M$ to denote all eventually-follows relations of the activity sequences allowed by a model $M$.

\noindent
For instance, $\mathit{EF}_{M_1} = \{ (\textit{receive order}, \textit{accept order}), (\textit{receive order}, \textit{reject order}),$ $(\textit{receive order}, \textit{deliver package}), (\textit{accept order}, \textit{deliver package})\}$

\section{Semantics-Aware Process Mining Tasks}
\label{sec:tasks}
This section describes and defines five process mining tasks that benefit from an understanding of process behavior. 
The tasks all focus on the control-flow perspective and are designed so that they do not involve (nor require) access to historical event data in order to perform them.
In this manner, we are able to assess whether a language model can solve the tasks based purely on its encoded knowledge of how processes generally work  with respect to the meaning of activities and their inter-relations.
Our tasks include two forms of \emph{semantic anomaly detection}, \emph{semantic next activity prediction}, and two flavors of \emph{semantic process discovery}. These tasks represent semantics-aware counterparts to well-established tasks in process mining and vary considerably in terms of their complexity.

\subsection{Semantic Anomaly Detection}
Anomaly detection in process mining focuses on identifying outlying process behaviors within the traces of an event log~\cite{van_der_aalst_process_2005}. Many approaches achieve this by detecting statistical outliers~\cite{bezerra_algorithms_2013}, based on the premise that infrequent behavior is anomalous.

In contrast, \textit{semantic} anomaly detection~\cite{van2021natural} focuses on identifying process behaviors that lack logical coherence. It challenges the assumption that infrequent behavior is necessarily anomalous and emphasizes that regular occurrences do not always constitute proper process behavior.
For example, from a semantic perspective, creating an invoice for a rejected purchase order constitutes anomalous behavior, regardless of how frequently it occurs.

Detecting anomalies based on process semantics requires a different approach compared to frequency-based anomaly detection.  
Whereas frequency-based detection can be performed by just using data in an event log (revealing statistical outliers), semantic anomaly detection requires information about how a process should (or should not) work in general. By definition, such information needs to be obtained from outside of the event log, e.g., from large knowledge bases or---as we do in this paper---from the knowledge encoded in LLMs.

We define two specific tasks in this context, focusing on the trace and activity-relation levels. 

\mypar{Trace-Level Semantic Anomaly Detection} 
Trace-level semantic anomaly detection (T-SAD) is a binary classification problem in which a trace $\sigma \in L$ needs to be classified as anomalous or not, based on its semantics and the set of possible activities $A_L$, which is provided as context information.
For instance, for a trace $\sigma=\langle$\emph{register application}, \emph{approve application}, \emph{review application}$\rangle$, the task is to classify that $\sigma$ is anomalous. This is because an application should first be reviewed and only then approved (or rejected).
The challenge here is that there is no specification available of the process at hand that can be used for this. Rather that anomaly needs to be inferred, requiring an understanding of how processes generally work.

\mypar{Activity-Level Semantic Anomaly Detection}
Activity-level semantic anomaly detection (A-SAD) is more fine-granular than T-SAD, focusing on pairs of activities in a trace rather than on an entire trace at once.
In particular, A-SAD focuses on classifying any eventually-follows relation $ a_i \prec_\sigma a_j$ of two activities $a_i$ and $a_j$ that appear in a trace $\sigma \in L$ as anomalous or not, based on its semantics and the set of possible activities $A_L$.
For instance, given the trace $\sigma=\langle$\emph{create purchase order}, \emph{reject purchase order}, \emph{create invoice}$\rangle$, the eventually-follows relation \emph{reject purchase order} $\prec_\sigma$ \emph{create invoice} should be classified as anomalous, whereas the other pairwise relations, i.e.,
\emph{create purchase order} $\prec_\sigma$ \emph{reject purchase order} and \emph{create purchase order} $\prec_\sigma$ \emph{create invoice} should be classified as valid.

\subsection{Semantic Next Activity Prediction}
Next activity prediction, also referred to as next event or next step prediction, is a fundamental task in predictive process monitoring~\cite{van2022process}. The goal is to predict the next activity in an ongoing process execution~\cite{neu2022systematic}. To address this task, numerous approaches, primarily leveraging supervised deep learning methods (e.g.,~\cite{evermann2017predicting, pfeiffer2021multivariate}), have been proposed.

As a semantics-aware counterpart for next activity prediction, we introduce the semantic next activity prediction (S-NAP) task. 
For an incomplete trace $\sigma$, which represents an ongoing process execution in which $k$ activities have been performed ($k\geq1$), the task is to predict the next activity $a_{k+1}$ in $\sigma \in L$ based on a set of possible activities $A_L$. 
For instance, given $\sigma = \langle$\emph{create purchase order}, \emph{approve purchase order}$\rangle$ and $A_L=\{$\emph{create purchase order}, \emph{approve purchase order}, \emph{create invoice}, \emph{make payment}$\}$, the task is to predict $a_{k+1}$ as \emph{create invoice}. This is because, generally, an invoice should be created before a payment is made. 

Whereas approaches for (traditional) next activity prediction train a model on historical traces from an event log $L$ to predict the next activity in ongoing (i.e., unseen) executions of the process, S-NAP focuses on situations where no such historical traces are available. As a result, the next activity must be inferred by considering the semantics of the activities involved in a process.

\subsection{Semantic Process Discovery}

Semantic process discovery focuses on generating a process representation that accurately captures the underlying semantics of a process, rather than reflecting observed behavior in event logs. 
Unlike traditional approaches that generate models based on behavioral relations recorded from historical process executions, semantic discovery incorporates domain knowledge to ensure that the resulting models represent proper and reasonable process behavior only based on a set of possible activities. 
This enables the detection of not what is frequent in event data but what is meaningful and correct in a process context. 
Such semantic models can form the basis for downstream tasks such as conformance checking, where we can compare the appropriate behavior (captured through the semantic model), to the actually observed traces.

For instance, a semantic discovery result should represent that a \emph{review application} activity must precede an \emph{approve application} activity, regardless of how frequently traces in the event log adhere to or violate this rule. This is because the process representation needs to align with logical process semantics rather than reflecting patterns in historical event data.
Discovering such representations requires knowledge about how processes should work, which can be sourced from domain-specific knowledge bases, human expertise or---as done in our work---LLMs.

We define two semantic discovery tasks, focusing on constructing directly-follows graphs and process trees:

\mypar{Semantic Directly-Follows Graph Discovery}
Semantic directly-follows graph discovery (S-DFD) is the task of generating a directly-follows graph that represents all reasonable relations between activities in a process. 
The goal is to produce a graph $D = (A, F)$, where each edge $(a,b) \in F$ reflects a valid directly-follows relation between activities $a$ and $b$,  only based on the set of activities $A$.
For instance, given a set of activities $\{$\emph{create purchase order}, \emph{approve purchase order}, \emph{reject purchase order}, \emph{create invoice}$\}$, the semantic DFG should include the edge (\emph{create purchase order}, \emph{approve purchase order}). This edge aligns with the common-sense understanding that an invoice is typically created following an approved purchase order.
Conversely, the DFG should not include the edge (\emph{reject purchase order}, \emph{create invoice}). 
This is because such a directly-follows relation contradicts the common-sense rule that an invoice should only be created for approved purchase orders.

The challenge lies in inferring the semantics of valid directly-follows relations without being able to rely on explicit process specifications and execution sequences.

\mypar{Semantic Process Tree Discovery}
Semantic process tree discovery (S-PTD) is a more structured task than S-DFD, focusing on constructing a hierarchical representation of a process, namely a process tree (see \autoref{sec:preliminaries}). The goal is to generate a process tree whose structure reflects the behavioral constraints in a process, such as parallelism, choices, and sequential behavior, only based on a set of possible activities $A$.
For instance, given the activity set
$\{$\emph{register application}, \emph{review application}, \emph{approve application}, \emph{reject application}$\}$, the semantic process tree should be $\rightarrow$(\emph{register application}, \emph{review application}, $\times$(\emph{approve application}, \emph{reject application})). 
This ensures that the applications are registered before they are reviewed and, finally, a decision is made, while only one outcome per review decision should be possible.

S-PTD is particularly challenging as it requires an understanding of not only pairwise relations but also how subsets of activities (parts of the entire process) relate to one another as well as how process trees work. Doing this without having historical execution data available,  necessitates external knowledge about process semantics to be able to construct trees that are both structurally and semantically sound.

\section{Datasets}
\label{sec:datasets}
This section outlines the creation and key characteristics of the corpus of process behavior and benchmarking datasets used to evaluate the ability of language models to solve the proposed tasks. All datasets are made publicly available~\cite{rebmann_2024_11276246}.

\subsection{A Corpus of Process Behaviors}
Language models require textual input. To assess their ability to solve semantics-aware process mining tasks, we need a collection of textual representations of process behavior, referred to as a corpus. This corpus serves as the foundation for creating task-specific data, which can then be used to train and evaluate language models on the proposed tasks.

\subsubsection{Corpus Creation}
As no suitable corpus is readily available, we create one based on graphical process models (i.e., process diagrams). For this purpose, we use \textsc{sap-sam}~\cite{sola2023sap}, the largest publicly available collection of process diagrams to date.
In order to create a high-quality corpus, we select only English BPMN diagrams from \textsc{sap-sam} that meet specific requirements. These ensure that the corpus includes only unique and valid process behavior.
In particular, we require that a diagram can be transformed into a sound block-structured workflow net and that no two diagrams have the same activity set. 
The former requirement mitigates data quality issues in the \textsc{sap-sam} collection~\cite{sola2023sap} and ensures that we can properly generate activity sequences from the diagram, and valid process trees from said activity sequences. 
The latter makes sure that different models in the corpus capture different process behavior and ensures a robust evaluation setup. Therefore, we discard a diagram, when another one with the same activity set is already present in the corpus.
Furthermore, we require a diagram to contain at least two different activities to ensure that it actually captures ordering relations between different activities.

We use the workflow net of each selected diagram to generate activity sequences, capturing all executions allowed by the net. 
For loops, we ensure that each loop is executed at most once, so that we capture relations involving rework, yet, obtain a finite set of activity sequences. 
For each net, this yields a process model $M$ according to the definition in \autoref{sec:preliminaries}, i.e., a set of activity sequences capturing its allowed behavior. 
We add each such $M$ to the corpus.

\subsubsection{Corpus Characteristics}
We show the characteristics of the resulting corpus in \autoref{tab:corpus}. As depicted there, the complexity of the process models varies considerably. For instance, the median number of unique activities is 4, whereas the maximum is 21 and the process models allow for 10.34 activity sequences on average, whereas the maximum amount is 10,080.

\begin{table}[!htb]
    \centering
    \caption{Characteristics of the process behavior corpus.}
    \label{tab:corpus}
    \begin{tabular}{lrrrrr}
    \toprule
     \textbf{Characteristic} & \textbf{Total} & \multicolumn{4}{c}{\textbf{Per process model}} \\
     & & \textbf{Avg.} & \textbf{Med.} & \textbf{Min.} &  \textbf{Max.}\\
    \midrule
    \textbf{\# Process models} &15,857 & -- & -- & -- & --\\
    \textbf{\# Unique activities} & 49,108 & 4.70 & 4 &2 &21\\
    \textbf{\# Unique sequences} &163,484 & 10.34 & 1 & 1& 10,080\\
    \bottomrule
    \end{tabular}
\end{table}

\begin{figure}[h]
    \centering
    \subfloat[Categorization by process type.\label{fig:apqc}]{%
        \includegraphics[height=6cm]{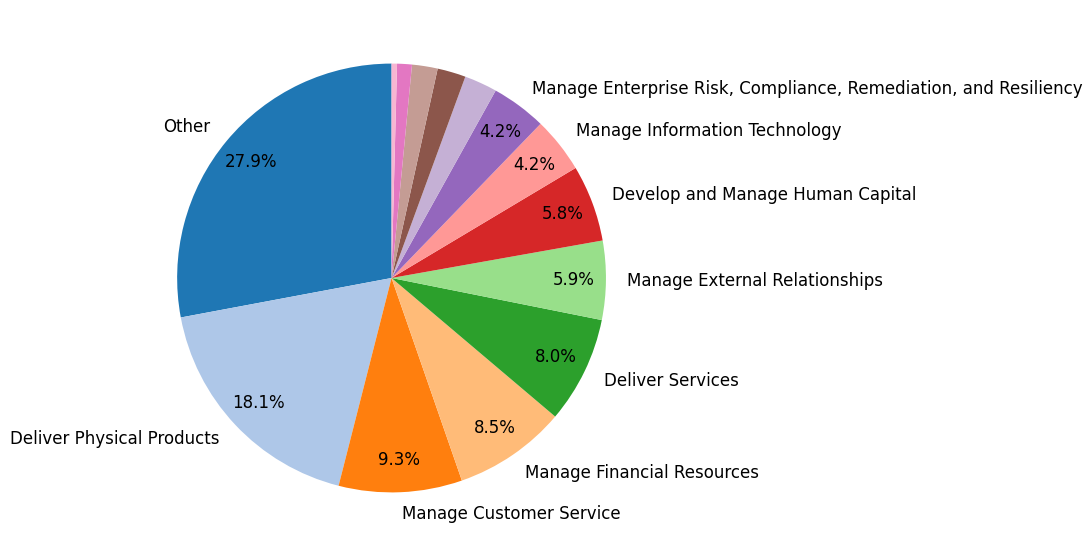}
    }
    \hfill
    \subfloat[Categorization by industry.\label{fig:sap}]{%
        \includegraphics[height=7.9cm]{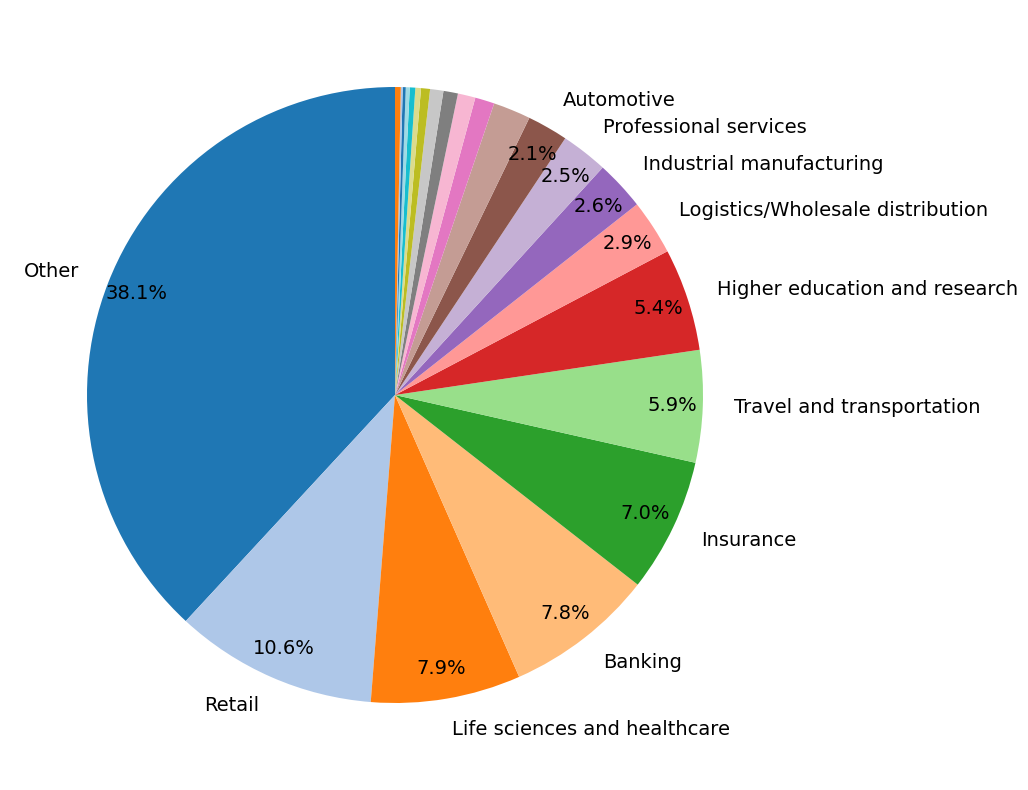}
    }
    \caption{Categorization of the process models in the corpus.}
    \label{fig:categorization}
\end{figure}

To highlight the broad coverage of the established corpus, we categorized its process models according to their process type and the industry they relate to, using established frameworks. 
We leveraged an LLM to perform the categorization itself. 
The LLM was prompted with each process model’s name, its unique set of activities, and the target APQC categories. Based on this, it assigned the most appropriate category or ``Other'' if no suitable match was found. 

For the categorization by \textit{cross-industry process types}, we used the first level of the APQC Process Classification Framework\footnote{\url{https://www.apqc.org/process-frameworks}}. 
The results, shown in \autoref{fig:apqc}, indicate that the corpus spans a wide range of process types, including HR, delivery of products and services, customer service management, and financial resource management. However, 28\% of the models could not be categorized using this framework.

For the \textit{industry-based categorization}, we applied the industry categories defined by SAP\footnote{\url{https://www.sap.com/industries.html}}. The resulting distribution, illustrated in \autoref{fig:sap}, shows broad representation across diverse industries, such as retail, healthcare, banking, insurance, and travel. Still, 38\% of the models could not be matched to any industry category.

Overall, while some models remained uncategorized per framework, only 14\% could not be assigned to a category in either of the frameworks. Manual inspection of these reveals that these cases largely consist of generic, low-level processes—such as downloading documents or filling in unspecified forms—as well as daily routine processes like preparing a meal or getting ready for work.

\subsection{Task-Specific Benchmarking Datasets}
Having established a text corpus of process behaviors, we are able to generate task-specific benchmarking datasets that include task samples and a gold standard. 
This gold standard enables objective, quantitative evaluation of language models on the tasks based on established evaluation measures such as F-$_1$ score for classification and fitness scores for discovery tasks.
The characteristics of the datasets are as follows:

\begin{itemize}
\item \emph{T-SAD} contains a total of 291,251 samples, of which 150,301 are valid, and 140,950 are labeled as anomalous. Trace lengths range from 1 to 17, with an average of 6.3 and a median of 7.
\item \emph{A-SAD} consists of 316,308 samples in total, equally divided between 158,154 valid samples and 158,154 anomalous samples.
\item \emph{S-NAP} comprises 1,289,081 samples in total, with prefix lengths ranging from of 1 to their full length. The mean prefix length is 4.6, and the median is 5.
\item \emph{S-DFD} consists of a total of 15,857 samples, with one sample per model. The number of edges ranges from a minimum of 1 to a maximum of 87, with an average of 5.2 and a median of 4.
\item \emph{S-PTD} features 15,857 samples as well (as for S-DFD there is one sample per original model).
\end{itemize}

\noindent We next describe how we established each of these datasets.

\subsubsection{Classification-Task Datasets}

\mypar{T-SAD} 
To establish the T-SAD dataset, we first create an event log $L$ for each process model $M$ in the corpus such that each $\pi \in M$ becomes a trace $\sigma \in L$. To make sure that there is a minimum number of traces per log, we randomly duplicate traces in $L$ until a size of 100 is reached if $L$ does not already contain at least 100 traces. 
Subsequently, for each trace $\sigma \in L$, we make a decision regarding the insertion of noise, with a 50 percent probability. 
This noise insertion involves swapping two randomly selected activities within the trace. 
After swapping, we check whether the resulting sequence $\sigma'$ is indeed anomalous, i.e., $\sigma' \notin M$. 
If the sequence is found to still be valid, i.e., $\sigma' \in M$, we continue iterating through potential swaps until we obtain an anomalous sequence\footnote{We limit the number of retries to 10 per trace to guarantee termination.}. 
This ensures that the dataset contains (roughly) the same amount of valid and anomalous traces, which is crucial for robust model training and evaluation. 

Each of the 291,251 records of the T-SAD dataset then consists of a trace $\sigma$, the correct label of $\sigma$, i.e.,  $\mathit{Anomalous}$ if $\sigma \notin M$ and $\mathit{Valid}$ otherwise, and the set of possible activities in the process from which $\sigma$ originates as context information.

\mypar{A-SAD} We create the A-SAD dataset based on the set of eventually-follows relations $\mathit{EF}_M$ for each process model $M$ in the corpus. 
The relations in $\mathit{EF_M}$ represent all valid execution orders of activities of $M$.
Next to these, we create a set of anomalous relations $\mathit{EF_{\not M}}$ , i.e., ones that are not in $\mathit{EF_M}$. 
To provide a balanced dataset, we establish $\mathit{EF_{\not M}}$ by randomly selecting relations that are not in $\mathit{EF_M}$, until we have an equal number of valid and anomalous relations.

Each of the 316,308 records of the A-SAD dataset consists of an eventually-follows relation $r$, the correct label of $r$, 
(i.e., if $r$ is anomalous or not), and the set of activities $A_M$ of the  process model from which $r$ originates (as context information).

\mypar{S-NAP} For the S-NAP dataset, we first create an event log $L$ for each process model $M$ in the corpus such that each $\pi \in M$ becomes a trace $\sigma \in L$.
Then, we generate all possible prefixes for each trace $\sigma \in L$ and add them to $L$. 
This involves iteratively considering sub-traces of increasing length $k$ from the first activity of a trace $\sigma$, up to one less than the full trace length, ensuring that there is always a subsequent activity available to serve as a prediction label.

Each of the 1,289,081 records of the S-NAP dataset then consists of a length-$k$ prefix ($\sigma_k$) of $\sigma$, the correct label of $\sigma_k$, i.e., the activity at position $k+1$ in $\sigma$, and the set of possible activities $A_L$ of the event log from which $\sigma$ originates as context information.

\subsubsection{Generation-Task Datasets}
\mypar{S-DFD} 
To establish the S-DFD dataset, we discover a DFG $D_M = (A_M, F_M)$ for each process model $M$ and its activity set $A_M$ following the definition in \autoref{sec:preliminaries}. The pairs in $F_M$ represent the valid directly-follows relation between two activities in the execution sequences of $M$.

As a result, each of the 15,857 records of the S-DFD dataset consists of the set of possible activities $A_M$ and the true DFG $D_M$, obtained from a  process model $M$.

\mypar{S-PTD}
We establish the S-DPT dataset based on the workflow nets we obtained during the creation of our process behavior corpus. For each workflow net that was used to generate a process model $M$, we translate it into a process tree $T_M$ according to the definition in \autoref{sec:preliminaries}.\footnote{This translation is straightforward as we only retained block-structured workflow nets when creating the corpus. For the details on how this translation exactly works, we refer to work on process discovery approaches, in particular, the Inductive Miner~\cite{leemans2013discovering}.}

Each of the resulting 15,857 records of the S-PTD dataset comprises the set of activities $A_M$ and the true process tree $T_M$, obtained from a process model $M$.

\section{LLM-Based Process Mining}
\label{sec:llms}
This section provides an overview of how we adapt (L)LMs for solving semantics-aware process mining tasks,  using few-shot in-context learning and fine-tuning for both the classification and generation tasks.

Neural language models, based on the Transformer architecture \cite{vaswani_attention_2017,devlin2019bert}, come in two main flavors: (1) bidirectional language models, also commonly called \textit{encoders}, which are typically (pre-)trained via masked language modeling objectives in which masked tokens are predicted from both left and right context \cite{devlin2019bert,liu2019roberta}, and (2) unidirectional language models, also known as \textit{decoders}, which are trained via autoregressive language modeling objectives where the next token is predicted from the preceding context \cite{brown2020language,touvron2023llama}. 
LLMs are large instances of the latter category (with at least a billion parameters) and are, following large-scale autoregressive language modeling, typically additionally trained for \textit{instruction following}, i.e., to provide solutions to tasks given the natural language description of these tasks \cite{wang2022super,zhang2023instruction}. 
Such instruction-tuning allows LLMs to generalize to new tasks through textual task descriptions (since capturing meaning of language is what LLMs excel at) and solve them successfully even when not provided with any or only few task-specific (training) examples (in-context learning).

\subsection{Few-Shot In-Context Learning}
In-context learning (ICL) aims to induce a model to perform a task by providing a small set of input-label examples (so-called ``shots'') along with the task description; the query sample---the example (input) for which the label is to be generated---is provided at the end of the \textit{prompt}~\cite{dong2022survey}. In-context learning allows the LLM to understand the task (via the description and a few labeled examples), without supervised fine-tuning (i.e., without any updates to the LLM's parameters).

\begin{figure}[!b]
    \centering
    \begin{subfigure}[t]{0.45\textwidth}
        \centering
        \includegraphics[width=\textwidth]{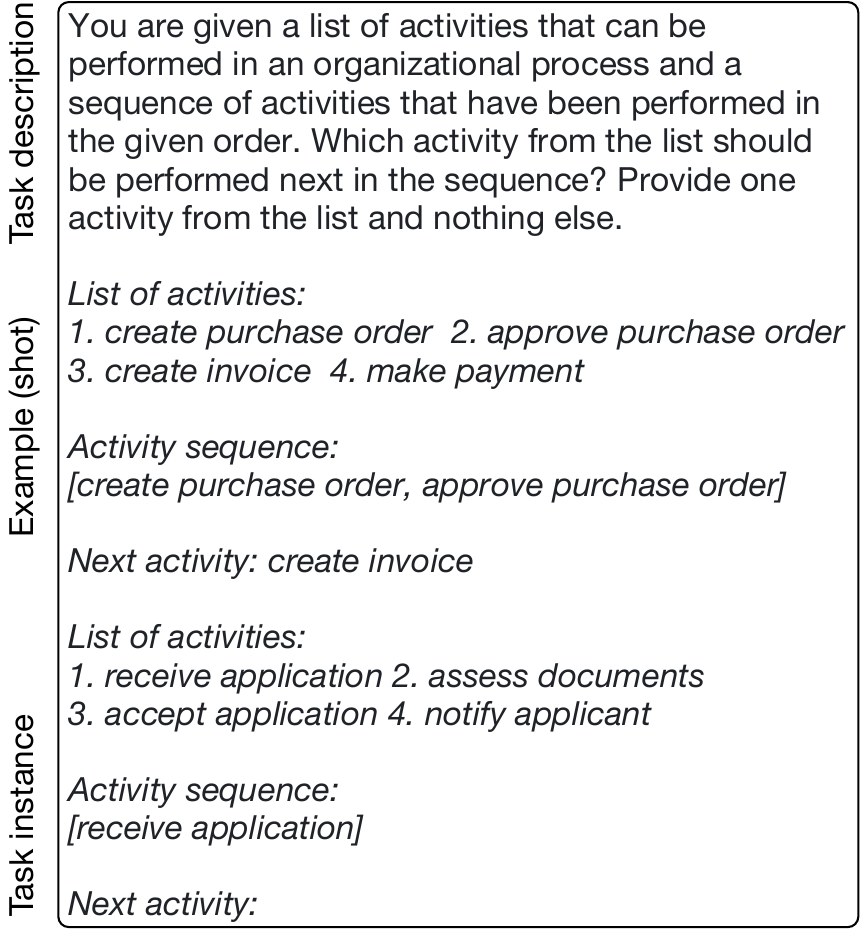}
        \caption{Prompt for the S-NAP task.}
        \label{fig:snap_prompt_a}
    \end{subfigure}
    \hfill
    \begin{subfigure}[t]{0.48\textwidth}
        \centering
        \includegraphics[width=\textwidth]{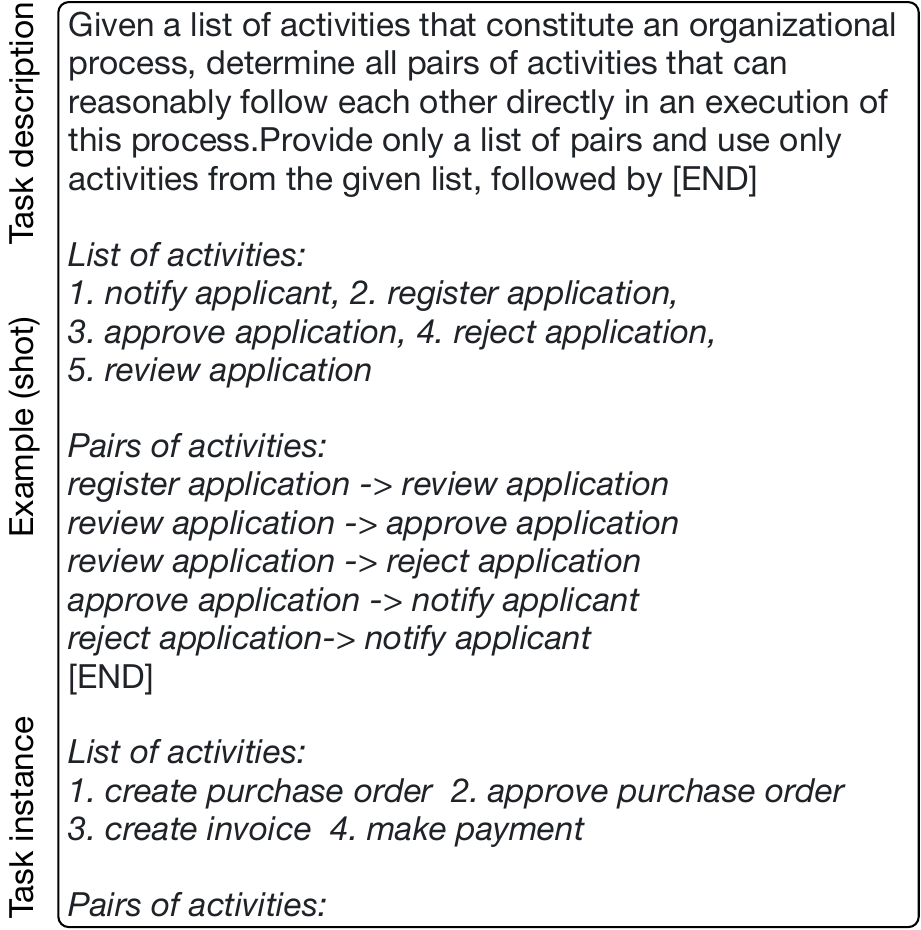}
        \caption{Prompt for the S-DFD task.}
        \label{fig:snap_prompt_b}
    \end{subfigure}
    \caption{One-shot in-context-learning prompts.}
    \label{fig:prompts}
\end{figure}

\autoref{fig:prompts} illustrates one-shot prompts for the S-NAP and S-DFD tasks. Each prompt begins with a description of the task, followed by a single labeled instance.
For the S-NAP task, the labeled instance includes a list of possible activities and the prefix trace in question, followed by the correct label. In contrast, the S-DFD task only presents the list of possible activities, followed by the corresponding true directly-follows pairs.
In both cases, the query instance for which the LLM is expected to generate a solution appears at the end of the prompt.

\subsection{Fine-Tuning LMs}
\label{subsec:ft-lm-clf}
Fine-tuning is the procedure of further training a pretrained language model, in order to specialize it for a specific task. 
The advantage compared to training a model from scratch is that the training data size for fine-tuning is considerably smaller, thus reducing resources required to train a task-specific model. 
We fine-tune (large) language models for each of our semantics-aware process mining tasks. While it is possible to train decoder LLMs for both classification and generation tasks we propose, the manner in which encoders are trained makes it difficult to adapt them for generation, as explained in the following sections.

\mypar{Fine-Tuning LMs on Classification Tasks}
Both encoder and decoder models can be tailored for classification tasks. We next describe the fine-tuning procedures for (a) discriminative classification with encoder language models and (b) generative classification with a decoder language models.    

\begin{figure}[!h]
    \centering
    \includegraphics[scale=0.63]{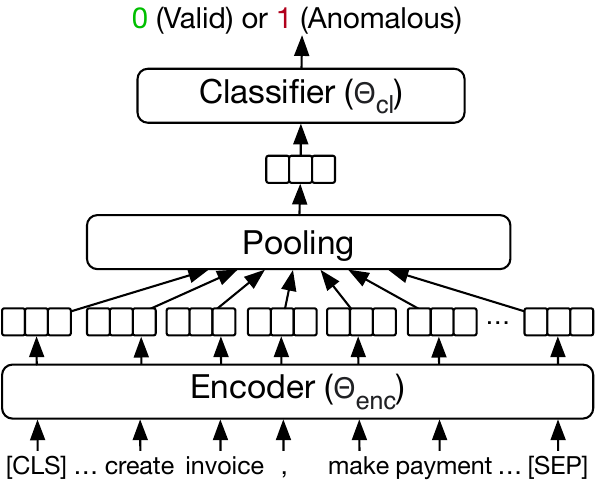}
    \caption{Illustration of discriminative classification with an \textit{encoder} LM.}
    \label{fig:encoder}
\end{figure}

\mypartwo{Discriminative Fine-Tuning of Encoder LMs}
To fine-tune an encoder LM for classification tasks, we extend the model's base architecture (i.e., the pretrained Transformer network) with an additional classification layer: the parameters of the classifier are trained from scratch (i.e., randomly initialized), whereas the encoder's parameters are updated (i.e., fine-tuned).
%%%
Fine-tuning of an encoder LM for the T-SAD task is illustrated in \autoref{fig:encoder} using a record with a trace $\langle$\emph{create invoice}, \emph{make payment}, $\dots \rangle$ as input. The input is first split into a sequence of subword tokens.\footnote{For more frequent words in the language, a token will commonly correspond to the whole word; less frequent words, on the other hand, will often be broken down into more frequent subtokens (e.g., \textit{``tokenization''} may be segmented into \textit{``token''} and \textit{``ization''}). The exact subword vocabulary is model dependent, i.e., each LM comes with its own tokenizer.} 
The actual input for encoder LMs is commonly surrounded with synthetic sequence start (\texttt{[CLS]}) and sequence end (\texttt{[SEP]}) tokens. 
The encoder (i.e., the Transformer network) outputs one vector---a transformed/contextualized representation---for each token in the input sequence, including the sequence start/end tokens. Let $\mathbf{x}_{CLS} \in \mathbb{R}^d$ be the representation of the sequence start token \texttt{CLS} (output of the encoder) with $d$ as the hidden size of the encoder's Transformer network; this vector $\mathbf{x}_{CLS}$ can be seen as a latent semantic representation of the whole input text and is forwarded as input to the classifier. The classifier, in turn, is a single-layer feed-forward network: $\hat{\textbf{y}} = \mathit{softmax}(\mathbf{W}_{cl} \cdot \mathbf{x}_{CLS} + \mathbf{b}_{cl})$; $\mathbf{W}_{cl} \in \mathbb{R}^{c \times d}$ and $\mathbf{b}_{cl} \in \mathbb{R}^{c}$ are the trainable parameters of the classifier ($c$ is the number of classes in the classification task) and $\mathit{softmax}$ is the function commonly used to convert real-valued vectors into probability distributions---the final output $\hat{\textbf{y}}$ is thus a probability distribution over the task's classes. We train the model (jointly update the parameters of both classifier and encoder in end-to-end fashion) by minimizing the widely used cross-entropy loss, i.e., the negative logarithm of the probability that the model predicted for the true class of the input instance.

T-SAD and A-SAD are binary classification tasks (i.e., $c = 2$) in which the model predicts whether the traces and ordered activity pairs, respectively, are \emph{Valid} or \emph{Anomalous}. S-NAP is a multi-class classification task in which the set of classes is defined by the activities in $A_M$ of the process model $M$ from which the input record was created.

\mypartwo{Generative Fine-Tuning of Decoder (L)LMs}
Autoregressively trained decoder LLMs cast classification tasks as language generation tasks. Concretely, each class into which the preceding text is to be classified is assigned one token from the vocabulary and the LLM's language modeling head (a classifier over the LLM's vocabulary) is supposed to generate the token of the correct class. 
For example, for the T-SAD task, we convert individual training instances into prompts that couple (1) the set of process activities with (2) the concrete trace (or activity sequence) that is to be judged as \textit{Valid} or \textit{Anomalous}. Then we append the prompt that asks whether the sequence is anomalous, with the token \textit{true} assigned to the \textit{Anomalous} sequences and token \textit{false} to the \textit{Valid} sequences. The whole input for the decoder LLM for a single sequence is shown below (the label token is underlined and in blue):   

\smallskip
\noindent \emph{Activities: \{create order, approve order, reject order, create invoice, make payment\}} \\
\emph{Activity sequence: [create order, reject  order, create invoice, make payment]}\\
\emph{Anomalous: \color{blue}{\underline{true}}}

\begin{figure}[!h]
    \centering
    \includegraphics[scale=0.63]{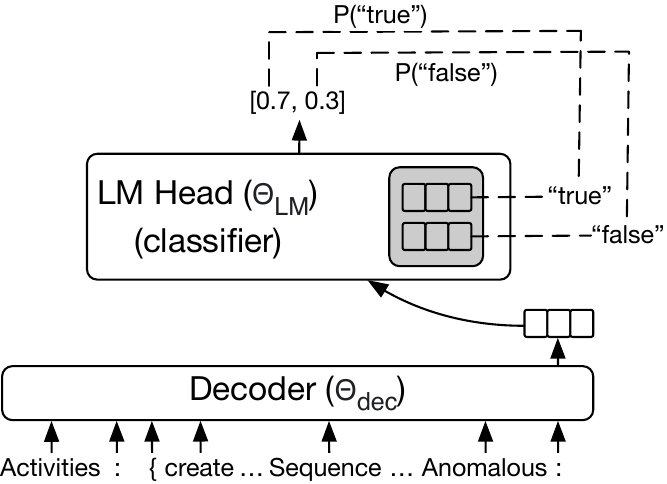}
    \caption{Illustration of constrained generative fine-tuning of a \textit{decoder} LM.}
    \label{fig:decoder}
\end{figure}

\smallskip
We fine-tune a decoder LLM via \textit{constrained text generation}: given the entire preceding context (everything except the last token that indicates the class), we predict the next token, but allow the language modeling head to only predict the probabilities for the \textit{allowed} class tokens (in the above example, only {\textit{true}, \textit{false}}), as opposed to LLM pretraining in which the next token is predicted over the entire vocabulary of the LLM. We illustrate constrained generative fine-tuning of a decoder LM in \autoref{fig:decoder}.     
The Transformer network of the decoder produces the output representation by contextualizing all preceding tokens; the resulting vector is next compared against the representations of the allowed class tokens (in the example, {\textit{true}, \textit{false}}) to produce scores that are then converted into probabilities using softmax. We minimize the negative log likelihood of the probability assigned to the correct class token: as updating all LLM parameters is computationally infeasible, we resort to parameter-efficient fine-tuning via low-rank adaptation (LoRA) \cite{hu2021lora}.\footnote{For brevity, we refer the reader to the original work for details on LoRA.}   

\mypar{Fine-Tuning LLMs on Generation Tasks}
In contrast to constraining text generation for classification tasks, for the generation tasks we allow the LLM to produce output freely, conditioned on the provided input context. 
This requires the LLM to develop a more holistic understanding of the S-DFD and S-PTD tasks. While for classification the outputs are limited to one of a set of predefined labels, the generation tasks involve creating complete answers. The model must account for what parts of the answer have already been generated and what remains to be completed.

The decoder LLMs we employ are pre-trained to solve tasks based on minimal context during instruction tuning. To adapt these models for our tasks, we fine-tune them on labeled instances comprising a textual description and a single task instance with the correct solution, i.e., a true DFG or process tree. 
This approach leverages both the extensive knowledge gained during pre-training and the instruction-following capabilities developed through post-hoc alignment. For example, for the S-PTD task, the input provided to the LLM for a single sequence includes the following components (with the label underlined and highlighted in blue):

\smallskip 
\noindent 
\emph{Given the list of activities that constitute an organizational process, determine the process tree of the process. A process tree is a hierarchical process model.} 

\noindent
[\textsc{description of the process tree notation}] 

\noindent
\emph{Make sure that you use each activity exactly once in the tree and that you set parentheses are correctly.}

\noindent 
\emph{Activities: {create order, approve order, reject order, create invoice}} 

\noindent 
\emph{Process Tree:} \underline{ \color{blue}{$\rightarrow$(\emph{create order}, $\times$(\emph{approve order}, \emph{reject order}), \emph{create invoice}) }} 

\medskip

As illustrated, we provide a detailed textual description of the task (as for ICL), followed by the list of activities for the specific task instance, and conclude with the expected output.

\begin{figure}[!h] 
\centering 
\includegraphics[scale=0.63]{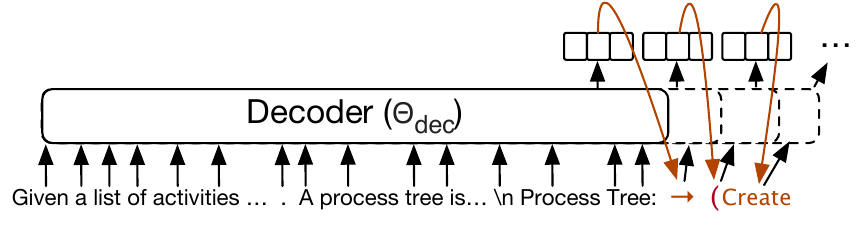} 
\caption{Illustration of generative fine-tuning of a decoder LM.} \label{fig:decoder_gen} 
\end{figure}

Importantly, the model is not trained to predict the next token for the provided context (the task description and the list of activities). Instead, it is fine-tuned to predict all the next tokens that belong to the \textit{label}, conditioned on the previous tokens in the correct order. The procedure is depicted in \autoref{fig:decoder_gen}. It ensures that the Transformer network generates output tokens iteratively, while taking into account the tokens it already produced as part of its answer. As for the classification tasks, we use LoRA to fine-tune LLMs for the generative tasks as well (cf. \autoref{subsec:ft-lm-clf}).

\section{Experimental Setup}
\label{sec:setup}
In this section, we first explain how the datasets are split into training, validation, and testing sets, followed by an introduction to the specific language models we use. We then detail our ICL and fine-tuning setups. To ensure reproducibility, we make all training and evaluation scripts publicly available.\footnote{\url{https://github.com/a-rebmann/llms4pm}; this repository also includes examples of how to use LLMs to solve individual instances of the five tasks.}

\mypar{Dataset Portions}
We split all datasets based on the process models from which the samples originate using 70\% of instances for training, 20\% for validation, and 10\% for final performance evaluation. 
For these splits, we ensure that no activity sequence from a model in the training set appears in any of the process models in the validation or test sets. This prevents any leakage of process behavior knowledge between the sets, allowing for an accurate assessment of the LMs' generalization abilities. Additionally, we use stratified sampling based on the number of unique activities in the models to ensure a comparable complexity distribution across the splits. The sizes of all splits for the five tasks are shown in \autoref{tab:split}.\footnote{Note that the distributions of process types and industries in the dataset portions are similar to the corpus. For the full details, we refer to our repository.}

\begin{table}[!htb]
    \centering
    \caption{Training, validation, and test split characteristics per task.}
    \label{tab:split}
    \begin{tabular}{lrrrr}
    \toprule
     \textbf{Task} & \textbf{Total} & \textbf{Train} & \textbf{Validation}& \textbf{Test} \\
    \midrule
    \textbf{T-SAD} &291,251 &227,892 &43,609&19,750 \\
    \textbf{A-SAD} & 316,308& 229,402& 56,154 & 30,752\\
    \textbf{S-NAP}& 1,289,081&1,071,529&166,811 & 50,741\\
    \midrule
    \textbf{S-DFD}& 15,780 & 11,397 & 2,775& 1,528\\
    \textbf{S-PTD}& 15,780 & 11,397 & 2,775& 1,528\\
    \bottomrule
    \end{tabular}
\end{table}

\mypar{Large Language Models}
We utilize two widely used open decoder-based LLMs that have demonstrated impressive performance on NLP benchmarks: (1) Llama-3 (\emph{Llama}) in its 8 billion parameter version\footnote{\url{https://huggingface.co/meta-llama/Meta-Llama-3-8B}} and (2) Mistral-2 (\emph{Mistral}) in its 7 billion parameter version\footnote{\url{https://huggingface.co/mistralai/Mistral-7B-Instruct-v0.2}}. We evaluate both models in few-shot ICL and fine-tuning setups.

\mypar{Baselines}
We use the following baselines to put the performance of LLMs on the proposed tasks into context:

\mypartwo{Classification}
For the classification tasks, we use three baselines:

\noindent
(1) For \emph{T-SAD} and \emph{A-SAD}, we compare LLMs with the state-of-the-art approach for rule-based semantic anomaly detection~\cite{van2021natural}. It is important to note that this baseline can only detect semantic anomalies if the respective activities refer to same business object. Consequently, it automatically classifies pairs of activities with distinct business objects as valid. We evaluate our approach against the best-performing configuration from the original paper, referred to as \emph{SEM4} in their experiments. 

\noindent
(2) We compare LLMs in an ICL setting (i.e., without task-specific fine-tuning) against a random classification baseline, which assigns one of the classes of the respective task to each test instance with equal probability per class. 

\noindent
(3) We compare generatively fine-tuned LLMs against a discriminatively fine-tuned encoder LM: specifically, we use \emph{RobERTa}~\cite{liu2019roberta} in its \emph{large} version,\footnote{\url{https://huggingface.co/FacebookAI/roberta-large}} a strong and widely used English-only bidirectional encoder LM. Note that fine-tuning \emph{RoBERTa} on \emph{A-SAD} corresponds to the state-of-the-art approach for detecting anomalous eventually-follows relations~\cite{caspary2023does}, with the difference being that for our experiments, we used a more powerful base model.

\mypartwo{Generation}
For the generation tasks, we established a baseline that randomly selects one of the behavioral footprint relations (defined in \autoref{sec:preliminaries}) for all pairs of activities in the respective set of possible activities. Note that the encoder baseline cannot be used for the generation tasks, due to the non-autoregressive pretraining and discriminative modeling objectives (cf.~\autoref{sec:llms}).

\noindent
Note that direct comparison with statistical baseline methods (e.g.,~\cite{bezerra_algorithms_2013} for anomaly detection) is not feasible since these tackle different tasks that take different inputs.  In particular, statistical approaches consider an entire event log, capturing data on numerous cases, in order to infer statistical patterns based on occurrence frequencies. Therefore, these approaches cannot be applied to the data that we have, where such frequencies are not available. Any comparison to statistical approaches would thus not be meaningful and will have a strong self-fulfilling character.

\mypar{Performance Measures} 
We use the following established performance measures to evaluate the classification and discovery tasks.

\mypartwo{Classification}
We measure classification performance using the macro F$_1$-score, so that classes equally contribute to the performance regardless of their size. Macro F$_1$ is the simple average of per-class F$_1$ scores, with F$_1$ of class $c$ being the harmonic mean between $c$'s precision and recall.

\mypartwo{Generation}
We measure generation quality with the well-known footprint-based fitness~\cite{carmona2018conformance}, which can be used to compare sets of allowed execution sequences based on the pairwise behavioral footprint relations introduced in \autoref{sec:preliminaries}.
In case of the S-DFD task, we establish the footprint of the gold standard DFG and the footprint of the LLM-generated DFG, according to the definition of a footprint in \autoref{sec:preliminaries}. Then, the footprint-based fitness is the fraction of equal footprint relations in both footprints.
For the S-PTD task, we first play-out the gold-standard and LLM-generated process trees to obtain sets of allowed execution sequences. We use these sets to establish DFGs and then proceed as for the S-DFD task to obtain the fitness score.

\mypar{In-Context Learning and Prompt Optimization}
We tested multiple task formulation prompts for each task and selected the one that yielded the best performance on the validation set. We evaluated 6-shot, 10-shot, and 20-shot ICL for the three classification tasks: for the binary T-SAD and A-SAD tasks, we evenly balance positive and negative instances; for S-NAP we sample one instance from randomly chosen 6, 10, or 20 process models, respectively. 
For the generation tasks, we evaluated 3-shot, 5-shot, and 10-shot ICL, which is required due to the longer prompt lengths compared to the classification tasks.
Different task description prompts led to marginal performance differences. Somewhat surprisingly, prompts with fewer shots produced better validation performance: thus, we finally evaluate 6-shot ICL on our test data for the classification and 5-shot for generation tasks.\footnote{See \url{https://github.com/a-rebmann/llms4pm} for the task prompts we used.}

\mypar{Fine-Tuning}
We fine-tune \emph{Llama} and \emph{Mistral} in batches of two instances with gradient accumulation over 16 batches, resulting in an effective batch size of 32. We fine-tune the \emph{RoBERTa} classification baseline in batches of 32 instances as well. 
All models are trained using the AdamW algorithm \cite{loshchilov2018decoupled}, with an initial learning rate of 1e-5. We fine-tune the LLMs for three epochs and \emph{RoBERTa} for ten epochs.  
We run each combination of task and model three times using different random seeds, corresponding to different random initialization of model parameters and shuffling of training data in each run.

\section{Results and Discussion}
\label{sec:results}
We first present the performance of LLMs on the classification tasks, followed by their performance on the generation tasks. Next, we provide an in-depth analysis of the language models' results, and conclude with a discussion of the training effort.

\subsection{Classification-Task Results}
\label{sec:results:single}
\autoref{tab:results:class} shows the results of our experiments on the classification tasks, along with the rule-based SAD baseline, the random classification baseline, the two LLMs (\emph{Llama} and \emph{Mistral}) using the ICL approach, as well as \emph{RoBERTa} and the LLMs using the fine-tuning approach (FT). We report mean macro $F_1$ scores as well as ($\pm$) standard deviation variance over three runs for ICL and three different random seeds for FT.

\begin{table}[!htb]
    \centering
    \caption{Classification Results (macro $F_1$ scores).}
    \label{tab:results:class}
\begin{tabular}{lccc}
\toprule
\multirow{2}{*}{\textbf{Approach}}& & \textbf{Task} & \\
 & \textbf{T-SAD}& \textbf{A-SAD} & \textbf{S-NAP}\\
 \midrule
 \textbf{Rule-based~\cite{van2021natural}} & 0.45 $\pm$ 0.000 & 0.33 $\pm$ 0.000 & - \\
\textbf{Random} & 0.50 $\pm$ 0.000 & 0.50 $\pm$ 0.000 & 0.13 $\pm$ 0.000 \\
\textbf{ICL Mistral} & 0.49 $\pm$ 0.022 & 0.44 $\pm$ 0.011 & 0.18 $\pm$ 0.018 \\
\textbf{ICL Llama} & 0.51 $\pm$ 0.015 & 0.53 $\pm$ 0.021 & 0.32 $\pm$ 0.054 \\ \midrule
\textbf{FT RoBERTa}  & 0.77 $\pm$ 0.006 & 0.85 $\pm$ 0.003 & 0.63 $\pm$ 0.048 \\ 
\textbf{FT Mistral} & \textbf{0.79} $\pm$ 0.010 & \textbf{0.88} $\pm$ 0.002 & 0.68 $\pm$ 0.039 \\
\textbf{FT Llama} & \textbf{0.79} $\pm$ 0.011 & \textbf{0.88} $\pm$ 0.000 & \textbf{0.69} $\pm$ 0.049 \\
\bottomrule
\end{tabular}
\end{table}

\subsubsection{In-Context Learning Results}
For ICL, we find that while the performance of the LLMs is consistently better than the rule-based baseline, it is at best marginally better (\emph{Llama}) at worst (\emph{Mistral}) slightly worse than random performance for the two semantic anomaly detection tasks, T-SAD and A-SAD. Specifically, \emph{Llama} achieves a macro $F_1$-score of 0.51 for T-SAD and 0.53 for A-SAD, while \emph{Mistral} scores 0.44 resp.\ 0.49. These results indicate that the LLMs have not effectively learned these tasks from the few examples provided in the context. For the S-NAP task, ICL with LLMs does outperform the random baseline, with \emph{Llama} exhibiting much stronger performance (19-point gain over the random baseline) than \emph{Mistral} (only 5-point gain). The performance is nonetheless fairly poor in absolute terms (mere 0.32 with \textit{Llama}).  
These results suggest that these process mining tasks drastically differ from the language processing tasks on which the LLM instruction-tuning was carried out. 

\subsubsection{Fine-Tuning Results}
Poor ICL performance shows that LLMs \textit{a priori} know very little about how to solve the proposed classification tasks and thus need to be explicitly trained for them. The fine-tuned encoder-LM baseline, i.e., \emph{RoBERTa}, already  achieves drastically better performance than ICL with LLMs: for example, it obtains the F$_1$-score of 0.77 on the T-SAD task, which is an improvement of massive 26 points over the best ICL performance (0.51 by \emph{Llama}).
Fine-tuning the LLMs yields even better performance, with \emph{Llama} and \emph{Mistral} achieving an $F_1$-score of 0.79, a further 2-point improvement over \emph{RoBERTa}'s performance. 
The same trend holds for the other two tasks: 
on A-SAD, both \emph{Mistral} and \emph{Llama} yield very strong performance $F_1$-scores of 0.88, outperforming \emph{RoBERTa} by 3 points; on S-NAP, \emph{Llama} and \emph{Mistral} achieve $F_1$-scores of 0.69 and 0.68, respectively (6- and 5-point respective gains over \emph{RoBERTa}).

These results show that decoder-based LLMs can effectively acquire the missing process knowledge through explicit task-specific fine-tuning, yielding better results than their (smaller) encoder-based counterparts such as \emph{RoBERTa}. 
The fine-tuned LLMs consistently outperform \emph{RoBERTa} on all proposed classification tasks, which points to the benefits of much larger-scale pretraining to which they have been comparatively exposed. 

\subsubsection{Task Comparison}
The results also demonstrate considerable differences in difficulty between the classification tasks. 
For A-SAD, the LLMs achieve an impressive $F_1$-score of 0.88, while the maximum score for T-SAD is 0.79, and the best model scores only 0.69 for S-NAP. This aligns with expectations. Solving T-SAD requires the model to identify whether process behavior is valid within the context of an entire trace, whereas A-SAD only requires assessing a single behavioral relation.
The S-NAP task is by far the most challenging of the classification tasks and simply unsolvable for many instances. 
For example, consider a process that allows for parallel execution of activities. Then, it is indeterminable---for both humans and automated approaches---which activity occurs next in a trace based solely on a prefix, as there are multiple valid options. 

\subsection{Generation-Task Results}
\autoref{tab:results:gen} shows the results for the generation-based process discovery tasks, along with the random generation baseline.
We report mean fitness scores and ($\pm$) standard deviation variance over three runs for ICL and three different random seeds for FT.

\begin{table}[!htb]
    \centering
    \caption{Generation Results (Fitness scores).}
    \label{tab:results:gen}
\begin{tabular}{lcc}
\toprule
\multirow{2}{*}{\textbf{Approach}}& \multicolumn{2}{c}{\textbf{Task}}  \\
 & \textbf{S-DFD}& \textbf{S-PTD} \\
\midrule
\textbf{Random} & 0.32 $\pm$ 0.008 & 0.32 $\pm$ 0.008 \\
\textbf{ICL Mistral} & 0.61 $\pm$ 0.008 & 0.56 $\pm$ 0.031 \\
\textbf{ICL Llama} & 0.60 $\pm$ 0.020 & 0.52 $\pm$ 0.044 \\ \midrule

\textbf{FT Mistral} & \textbf{0.81} $\pm$ 0.002 & \textbf{0.84} $\pm$ 0.004 \\
\textbf{FT Llama} & 0.80 $\pm$ 0.001 & 0.83 $\pm$ 0.015  \\
\bottomrule
\end{tabular}
\end{table}

\subsubsection{In-Context Learning Results}
The ICL results for the discovery tasks reveal considerable improvements over generating random behavioral relations for both the S-DFD and S-SPT variations. 
For example, while the random baseline achieves a fitness score of 0.32, \emph{Mistral} attains a score of 0.61 for S-DFD, representing an improvement of nearly 100\%. 
For S-PTD, the performance gains are slightly smaller, with \emph{Mistral} achieving a 0.24 increase and \emph{Llama} a 0.2 increase. Interestingly, \emph{Mistral} outperforms \emph{Llama} on the generation tasks, whereas \emph{Llama} performed better on classification tasks.

Overall, these findings suggest that the process knowledge encoded in LLMs during pretraining enhances their ability to address semantic discovery tasks compared to the baseline. 
The weaker ICL results on the classification tasks further indicate that LLMs grasp the process-related generation tasks better when provided with a handful of examples compared to the classification tasks. This difference likely stems from the closer resemblance of the generation tasks to the types of tasks the LLMs encountered during pretraining. Although performance gains compared to the random baseline are substantial, a fitness value of 0.6 on average still leaves room for improvement, suggesting that fine-tuning is also required for semantic process discovery tasks.

\subsubsection{Fine-Tuning Results}
Fine-tuning the LLMs on the generation tasks yields considerable performance gains, similar as for the classification tasks. In particular, \emph{Mistral} and \emph{Llama} achieve average fitness scores of 0.81 resp. 0.8 for S-DFD and 0.84 resp. 0.83 for S-DPT. These are massive improvements of up to 20 points for S-DFD and even 28 points for S-PTD compared to the best ICL performance. When looking at the stability of the results across evaluation runs, we find that on average results vary only marginally, with a standard deviation of less than 0.02 for both tasks and LLMs. 
Given that different models may be possible for a given set of activities, we also assess the stability of results for the same activity sets across the three evaluation runs. The results indicate that in more than 60\% of the cases the results are stable, leading to a standard deviation of 0 across runs. 
However, in the remaining cases, we observe a positive standard deviation between 0.007 and 0.43, with a median of 0.12. This indicates the presence of ambiguous activity sets in relation to the resulting model.
When examining the results in detail, we find that deviations of more than 0.25 primarily occur for small activity sets of up to four activities. 
These sets typically correspond to relatively sequential processes, yet the LLMs generate models that allow for numerous interleaving relationships between them. Thus, the ambiguity with respect to the resulting model has a more significant impact on fitness in smaller activity sets, which is in line with expectations.\footnote{We provide the raw results of this analysis in our repository.}

These results underscore the advantages of fine-tuning LLMs also for process-related generation tasks. The notable performance improvement in the process tree discovery task demonstrates that fine-tuning is especially beneficial for tasks with multiple facets. In particular, solving S-PTD requires not only a grasp of process semantics but also an understanding of a process-specific representation format. LLMs can effectively acquire both capabilities through targeted task-specific fine-tuning.

\subsubsection{Task Comparison}
The ICL results show that the LLMs can solve the S-DFD task better than S-PTD, which is in line with expectations, given that the discovery of DFGs is less complex than the discovery of process trees. 
For instance \emph{Llama} achieves 0.6 fitness on average for S-DFD, yet, only 0.52 for S-PTD. This is in line with expectations, since the former requires an LLM to grasp only pairwise relations, whereas the latter requires understanding how entire process parts relate to each other, while, in addition, the representation format is not trivial.

However, after fine-tuning the LLMs, this trend flips as shown by the performance for the S-PTD task, which exceeds the performance on S-DFD. 
For example, \emph{Llama} achieves an average fitness score of 0.8 for S-DFD but an impressive 0.83 for S-PTD.
This not only shows that complex process-related generation tasks can be effectively learned by LLMs, but also suggests that, by means of fine-tuning, LLMs can learn the relatively few global behavioral relations in a process more effectively than the many local pair-wise relations.

\subsection{In-Depth Analysis}
\label{sec:results:indepth}
To assess whether the fine-tuned language models handle certain processes of different types and industries more effectively than others, we conducted an in-depth analysis of their performance. Our primary focus is on S-PTD, the most intricate semantics-aware process mining task, where we provide a detailed breakdown across the different process types and industries. To ensure comprehensive coverage, we also include an analysis of T-SAD---as a representative classification task---offering detailed insights into model performance for this task category as well.

\subsubsection{In-Depth Analysis of \emph{S-PTD}}
We aim to understand whether performance generally differs between process types and industries. To this end, we report on the results of \emph{FT Mistral} for S-PTD, which align with those for \emph{FT Llama}. \autoref{tab:process_fitness} shows the results per process type and \autoref{tab:industry_fitness}  per industry.

\begin{table}[htbp]
\centering
\caption{Process type fitness scores of \emph{FT Mistral} for S-PTD and the share of the type in the test set (min. 0.5\%).}
\label{tab:process_fitness}
\begin{tabular}{lcc}
\toprule
\textbf{Process Type} & \textbf{Fitness} & \textbf{Share (\%)} \\
\midrule
Manage Enterprise Risk, Compliance, \dots & 0.89 & 3.80 \\
Market and Sell Products and Services & 0.86 & 2.53 \\
Manage External Relationships & 0.86 & 6.33 \\
Deliver Physical Products & 0.86 & 16.79 \\
Deliver Services & 0.85 & 8.10 \\
Manage Customer Service & 0.85 & 8.95 \\
Other & 0.82 & 27.59 \\
Manage Financial Resources & 0.82 & 10.55 \\
Develop and Manage Business Capabilities & 0.80 & 1.77 \\
Acquire, Construct, and Manage Assets & 0.78 & 1.77 \\
Manage Information Technology & 0.78 & 4.05 \\
Develop and Manage Human Capital & 0.77 & 5.49 \\
Develop and Manage Products and Services & 0.73 & 1.86 \\
\bottomrule
\end{tabular}
\end{table}

We find that for \emph{Deliver Physical Products}, with a substantial share of 17\%, the model achieved a solid fitness score of 0.86. In contrast, \emph{Manage Financial Resources} saw a lower score of 0.82, indicating relatively weaker model performance on more abstract financial processes compared to tangible operational ones. The model also appears to struggle with more knowledge-intensive processes, such as \emph{Develop and Manage Products and Services}, scoring below 0.75. Overall, there is no clear correlation between a process type’s share in the test set and its fitness score, though.

\begin{table}[htbp]
\centering
\caption{Industry fitness scores of \emph{FT Mistral} for S-PTD and the share of the industry in the test set (min. 0.5\%).}
\label{tab:industry_fitness}
\begin{tabular}{lcc@{\hspace{1cm}}lcc}
\toprule
\textbf{Industry} & \textbf{Fitness} & \textbf{Share (\%)} & \textbf{Industry} & \textbf{Fitness} & \textbf{Share (\%)} \\
\midrule
Insurance & 0.93 & 7.00&  Other & 0.81 & 37.64  \\
Media, \dots & 0.90 & 1.01 & Healthcare & 0.81 & 9.11 \\
Retail & 0.88 & 9.03 &  Pro. services & 0.81 & 2.11\\
Retail & 0.88 & 9.03 &  Automotive & 0.81 & 2.53\\
Logistics, \dots & 0.87 & 2.19 &  Edu., Research & 0.80 & 5.40  \\
Travel & 0.85 & 5.06 &  Cons. products & 0.78 & 1.52 \\
High tech & 0.85 & 0.68 & Government & 0.73 & 1.01 \\
Banking & 0.84 & 9.11           &  Real estate, \dots & 0.66 & 0.76\\
Indust. manuf. & 0.82 & 3.80     &  &  &  \\
\bottomrule
\end{tabular}
\end{table}

Among the industries, \emph{Retail} and \emph{Banking}, with shares of 9.03\% and 9.11\% respectively, achieved good fitness scores of 0.88 and 0.84, indicating solid model performance on processes of these industries. In contrast, lower scores were observed for industries such as \emph{Government} (0.73), suggesting difficulties in handling sectors with potentially more complex or specialized processes, which are also underrepresented in both training and test data. Interestingly, the model also struggled with knowledge-driven industries such as \emph{Education and Research} (0.80) and \emph{Professional Services} (0.81).

Looking at individual task instances, we find that in many cases, the LLM-generated process trees allow for the same execution sequences as the true process trees, which demonstrates the LLMs' ability to generate semantic process trees.

For instance, for the true tree $\rightarrow$(\emph{Complete purchase request}, \emph{Send to clerk}, \emph{Enter into system}, \emph{Fax PO}, \emph{Ship Product}, \emph{Receive Shipment}), \emph{Mistral} generated an identical tree, yielding a fitness of 1. 
But also beyond purely sequential processes in standard domains such as purchasing, the LLMs produce perfect outcomes. Examples of this include the grocery checkout process $\rightarrow$(\emph{Scan Items}, \emph{Scan Rewards Card and request payment}, $\times$(\emph{Process Credit Card}, \emph{Accept Cash}), \emph{Bag Grocery items}) and the risk assessment process $\rightarrow$(\emph{Risk threshold assessment}, $\times$(\emph{Advanced risk assessment}, \emph{Simple risk assessment}), \emph{Passed assessment}, $\wedge$(\emph{Notify customer with result}, \emph{Organize disbursement})). In both cases, only syntactical differences were present (such as inverse ordering in exclusive or parallel structures).

Even for more challenging cases such as a prescription-fulfillment process, the LLM achieves high fitness scores (here 0.85). In this case the true tree, $\rightarrow$($\times$(\emph{Collect walk-in prescription}, $\rightarrow$(\emph{Drop prescription in the appropriate box}, \emph{Pick up prescriptions in the box})), \emph{Enter prescription details}, \emph{Validate prescription}, \emph{Check insurance coverage}, \emph{Fill prescription}, \emph{Prescription pick-up request}), only differs from the generated tree, $\rightarrow$($\times$(\emph{Collect walk-in prescription}, $\rightarrow$(\emph{Drop prescription in the appropriate box}, \emph{Pick up prescriptions in the box}), $\rightarrow$(\emph{Enter prescription details}, \emph{Validate prescription}, \emph{Fill prescription})), \emph{Check insurance coverage}, \emph{Prescription pick-up request}), in that insurance coverage must be checked before the prescription is filled in the true tree, whereas the inverse is the case in the generated tree.

There are also cases where the LLM generates process trees that diverge from the true tree. For instance, while the true tree $\rightarrow$(\emph{Process trip information}, \emph{Check credit card information}, \emph{Process request}, \emph{Notify customer}) is purely sequential, the LLM generated a tree with a parallel construct ($\wedge$), implying that \emph{Process trip information} and \emph{Check credit card information} can happen in any order. 
This is one on many instances where the generated tree arguably represents a reasonable process, yet, only yields a low fitness score (in this case 0.625). Other examples include a travel-reconciliation process with a fitness score of only 0.2. 
According to the true tree, $\wedge$(\emph{Verify accounts}, \emph{Verify payment information}, \emph{Archive the form}, \emph{Authorize the travel-advance-reconciliation form}, \emph{Accept payment}), all activities can be performed in any order, which is clearly problematic. For instance, the verification of payment information should precede the acceptance of payment. In contrast, the generated tree arranges the process in a purely sequential order: $\rightarrow$(\emph{Verify Accounts}, \emph{Verify Payment Information}, \emph{Accept Payment}, \emph{Authorize the Travel-Advance-Reconciliation Form}, \emph{Archive the Form}). Although this sequential arrangement is not ideal either, both this and the previous example illustrate that the process models in the employed collection do not always represent semantically correct models\footnote{We aim to mitigate this issue in the future by establishing additional evaluation data based on a quality-assured model collection (see also \autoref{sec:conclusion}).}, and that there are cases where multiple acceptable solutions exist.

\subsubsection{In-Depth Analysis of \emph{T-SAD}}
For the T-SAD classification task, we find that the LLMs accurately detect anomalous traces across a wide range of process types and industries as well. For this classification task, we also relate the performance to encoder baseline, \emph{RoBERTa}.
Both the LLM and \emph{RoBERTa} appear to be particularly effective in identifying anomalies for standard process types. For example, in a claim-handling process, they correctly identify anomalies such as $\langle$\emph{Enter and verify claim}, \emph{Handle payment}, \emph{Assess claim}$\rangle$, where the claim should be assessed before sending a payment. They also correctly detect that the trace $\langle$\emph{Confirm order}, \emph{Ship product}, \emph{Get shipment address}, \emph{Emit Invoice}, \emph{Receive Payment}$\rangle$ is anomalous given that the product is shipped before the address is determined in this order-handling process. 

For more specialized industries such as healthcare, LLMs often outperform \emph{RoBERTa}. For instance, \emph{Llama} correctly identifies the trace $\langle$\emph{Arrival}, \emph{Treatment}, \emph{Triage}, \emph{Discharge}, \emph{Invoicing}$\rangle$ of a hospital process as anomalous, since \emph{Triage} should occur before \emph{Treatment}. In contrast,  \emph{RoBERTa} fails to detect this anomaly. Conversely, \emph{RoBERTa} incorrectly flags the trace $\langle$\emph{Disassemble system}, \emph{Refurbish materials}, \emph{Clean and paint covers}, \emph{Mount materials}, \emph{Move to bay}, \emph{Calibrate}, \emph{Handover}$\rangle$ of a refurbishing process as anomalous, even though it is valid, whereas the fine-tuned LLM correctly classifies this trace as valid. 

Finally, there are instances where both the LLMs and \emph{RoBERTa} incorrectly identify valid traces as anomalous. For example, both LLM and \emph{RoBERTa} flag the trace $\langle$\emph{Receive invoices of partners}, \emph{Handle payment of customer}, \emph{Receive review}, \emph{Send payment to partners}$\rangle$ as anomalous, even though it is valid. According to the corresponding process model in the corpus, a review can be received at any point during an execution of the process, making this trace valid. However, this specificity might also be challenging for a human to determine without further contextual information. 

\subsection{Training Effort}
\label{sec:results:effort}
Finally, we consider the effort required to train the language models on the tasks. 
Since in-context learning does not require any training effort, as the task-specific knowledge is provided at inference time, we focus on the training effort of fine-tuning an LLMs. For the classification tasks, we can also put this effort into context by comparing against fine-tuning an encoder (\emph{RoBERTa}).

\autoref{tab:results:effort} shows the run times for fine-tuning the LLMs (\emph{Llama} and \emph{Mistral}) for the different tasks per epoch, i.e., pass over all training samples.
As shown, the LLMs require a considerable amount of time for training across tasks. 
The training time varies between three hours per epoch for the semantic process tree discovery task and up to 23 hours for the next activity prediction task. 
This variance is predominantly caused by the considerably different number of training samples per task, which ranges from roughly 16 thousand for the generation tasks and 1.3 million for the next activity prediction task. 

When comparing the training effort for \emph{Llama} on the classification tasks with the encoder baseline, we observe substantial differences. In particular, training \emph{Llama} on the tasks takes up to 25 times longer than training \emph{RoBERTa} per epoch on the same data.
For instance, while fine-tuning \emph{Llama} for A-SAD takes 15 hours, \emph{RoBERTa} requires only around 40 minutes per epoch. This difference can be attributed to the huge number of parameters that need to be updated for the LLM during fine-tuning, even when using parameter-efficient fine-tuning. However, it is worth stressing that the LLM requires considerably fewer epochs to converge in terms of validation loss across tasks. This indicates that it not only learns the tasks better (as shown in the previous subsections), but also with fewer passes over the training data. 

\begin{table}[!htb]
    \centering
    \caption{Average run times for fine-tuning (per epoch).}
    \label{tab:results:effort}
    \begin{tabular}{lrrrrr}
\toprule
\multirow{2}{*}{\textbf{Approach}}& \multicolumn{5}{c}{\textbf{Task}} \\
 & \textbf{T-SAD} & \textbf{A-SAD} & \textbf{S-NAP} & \textbf{S-DFD} & \textbf{S-PTD}  \\
\midrule

\textbf{FT Mistral} & 9.9h & 14.3h & 21.9h & 4.6h & 3.3h \\
\textbf{FT Llama} & 11.1h & 15.0h & 23.0h & 4.2h & 3.0h \\
\midrule
\textbf{FT RoBERTa} & 0.5h & 0.6h & 1.3h & - & -  \\
\bottomrule
\end{tabular}
\end{table}

\section{Related Work}
\label{sec:related}
The natural language understanding capabilities of neural language models have been applied to various process analysis tasks, such as, extracting process information from text~\cite{bellan2022extracting, bellan2021process}, annotating event logs with semantic information~\cite{rebmann_enabling_2022}, detecting semantic anomalies~\cite{caspary2023does, busch2024xsemad}, and generating event logs from textual records of process steps~\cite{kecht2021event}.

%An exception is the work by Busch et al.~\cite{busch2024xsemad}, which investigates the ability of a  sequence-to-sequence language model to correctly instantiate declarative process constraints based on a given set of activities. Given the creation of a gold standard, the work can be assessed using established evaluation metrics. Furthermore, this task aligns closely with our S-DFD task, as it focuses on identifying specific types of behavioral relations between activities.
\mypar{Application of LLMs for Process Analysis}
With the advent of LLMs, researchers have increasingly explored their potential in process analysis~\cite{torres2024mapping}. Key applications include transforming textual process descriptions into formal process models~\cite{grohs2023large, kourani2024process, klievtsova2023conversational, ziche2024llm4pm, nour2024decomposed}, generating textual descriptions from process data such as models and event logs~\cite{berti2023abstractions}, identifying bottlenecks or undesired process behaviors~\cite{berti2023abstractions}, and abstracting fine-granular events into higher-level ones~\cite{fani2023llms}. 
However, much of this research has been conducted using closed-source GPT models or proprietary software products like \emph{ChatGPT}, which limits the ability to perform structured and reproducible evaluations~\cite{torres2024mapping}.

\mypar{Evaluation of LLMs on Process Analysis Tasks}
The lack of rigorous evaluations of LLMs for process mining tasks has recently been highlighted within the process mining community~\cite{torres2024mapping, berti2024evaluating}. In response, Berti et al.\ introduced a benchmark for process mining analysis questions~\cite{berti2024pm}. This benchmark comprises 52 prompts used to query various LLMs, with their responses evaluated by a closed-source GPT model. Although the benchmark provides interesting insights, using an LLM to rate the results can yield biased outcomes. Such bias arises, e.g., from the tendency of LLMs to favor their own output~\cite{panickssery2024llm}. 
Beyond this process-mining-specific effort, Busch and Leopold~\cite{busch2024towards} present a LLM-benchmark for a set of business process management tasks, including the recommendation of activities during the modeling of processes and the identification candidates for robotic process automation. Focusing on process modeling, Kourani et al.~\cite{kourani2024evaluating} present a benchmark that evaluates various LLMs using a set of 20 curated business processes.

In contrast to these works, we define process mining tasks that benefit from an understanding of process behavior. 
Furthermore, we evaluate LLMs using extensive task-specific benchmarking datasets in both, in-context learning and fine-tuning settings. The latter has not been investigated by other works, yet, as our experiments show, can be highly beneficial for challenging process-related tasks. 
Finally, it is important to note that our datasets provide gold standards, which allows for using established evaluation measures for classification tasks, eliminating the need for a proxy LLM to assess output quality.

\section{Conclusion}
\label{sec:conclusion}
In this paper, we investigated the capabilities of LLMs to solve semantics-aware process mining tasks, i.e., tasks that benefit from an understanding of the meaning of process steps and their relationships. 
We defined five such tasks and provide an extensive benchmarking dataset for each of them.
Our evaluation experiments that use these datasets show that LLMs fail to solve the tasks in in-context learning settings. 
However, our results demonstrate that LLMs achieve accurate performance when fine-tuned for these tasks. 
Furthermore, they surpass smaller, encoder-based language models in both scope (the types of tasks they can solve) and accuracy (the quality of how they solve tasks).

In the future, we want to investigate the integration of state-of-the-art process mining approaches with LLMs. 
Since we have shown that LLMs can solve semantics-aware process mining tasks through encoded knowledge of process semantics, integrating existing process mining approaches with LLMs may yield performance improvements for classical process mining tasks they address. 
For example, an existing next-activity prediction approach could be extended by an LLM-based semantic check that rejects predictions that do not make sense, thereby improving the overall prediction performance.
Furthermore, we plan to perform evaluation experiments on additional data. In particular, we want to generate a process behavior corpus and benchmarking data sets based on a large cross-domain collection of real-world reference process models, in order to investigate the generalizability of our results beyond the academic process model collection that we employed in this work. 
Finally, we aim to take first steps towards \emph{process fine-tuning}, i.e., creating LLMs that are specialized for handling not one, but many process analysis tasks. More concretely, instead of fine-tuning an LLM for a specific semantics-aware process mining task that it can then solve better, our goal is to fine-tune a model on task-solution pairs of many process analysis tasks, improving its performance on process-related tasks in general, such as, for instance, envisioned as part of a \emph{Large Process Model}~\cite{kampik2024large}.

\section*{Declarations}
 
\mypar{Ethical Approval} 
Not applicable.
 
\mypar{Funding} 
Not applicable.
 
\mypar{Availability of data and materials} 
Our training and evaluation scripts are accessible via the project repository linked in \autoref{sec:setup}. The process behavior corpus and benchmarking datasets are published separately~\cite{rebmann_2024_11276246}.

\mypar{Competing Interests}
The corresponding author works for a company that builds process mining software.

\mypar{Acknowledgment}
Our research was supported by the state of Baden-Württemberg through bwHPC.

\bibliography{sn-bibliography}% common bib file

\end{document}